\documentclass[epj,twocolumn]{webofc}
\usepackage[varg]{txfonts}   % Web of Conferences font
\woctitle{ISVHECRI 2016}
\begin{document}
\title{Parametric Analysis of Cherenkov Light LDF from EAS for High Energy Gamma Rays and Nuclei: Ways of Practical Application}

\author{A.Sh.M. Elshoukrofy\inst{1,2}\fnsep\thanks{\email{abeershehatamahmoud@yahoo.com}} \and
        E.B. Postnikov\inst{1}\fnsep\thanks{\email{evgeny.post@gmail.com}} \and
        E.E. Korosteleva\inst{1} \and
        L.G. Sveshnikova\inst{1}\fnsep\thanks{\email{tfl10@mail.ru}} \and
        H.A. Motaweh\inst{2}
        % etc.
}

\institute{Lomonosov Moscow State University Skobeltsyn Institute of Nuclear Physics (MSU SINP), 1(2) Leninskie gory, GSP-1, 119991, Moscow, Russia
\and
           Faculty of Science, Damanhour University, El-Gomhouria St., 22516, Damanhour, El Beheria, Egypt
          }

\abstract{%
In this paper we propose a 'knee-like' approximation of the lateral distribution of the Cherenkov light from extensive air showers in the energy range 30-3000 TeV and study a possibility of its practical application in high energy ground-based gamma-ray astronomy experiments (in particular, in TAIGA-HiSCORE). The approximation has a very good accuracy for individual showers and can be easily simplified for practical application in the HiSCORE wide angle timing array in the condition of a limited number of triggered stations. }
\maketitle
\section{Introduction}
\label{intro}

Ground-based gamma-ray astronomy has developed very fast since the discovery of the first TeV gamma-ray source, the Crab Nebula, in 1989 \cite{1} by the Whipple collaboration. This experiment was based on the imaging air Cherenkov technique, IACT \cite{2}. The main idea of this technique is the use of a telescope for collecting the Cherenkov light produced in the atmosphere by very energetic charged particles from extensive air showers (EAS). The competing method of Cherenkov light registration for gamma-ray astronomy is a wide angle timing array, which was first realized in THEMISTOCLE \cite{3} and AIROBICC \cite{4}. At the  present time the gamma-ray timing array concept HiSCORE \cite{5, 6} (High Sensitivity Cosmic Origin Explorer) is realized in the Tunka valley in Siberia as  part of the TAIGA observatory \cite{7}. This is an array of wide-angle non-imaging Cherenkov light detectors spaced about hundred meters apart from each other. The wide angle technique assumes that the air shower front timing is measured by counting the number of Cherenkov photons, $Q$, emitted by secondary air shower particles and subsequently registered in every $i$-th station of the array at time, $t$, at a distance, $R$, from the shower core $Q_i(R, t)$. The analysis of $Q_i(R, t)$ gives the possibility to reconstruct all the parameters of the primary particle needed for gamma-ray astronomy: shower arrival direction, core position, energy and type of particles \cite{8, 9}.

The critical point of the reconstruction procedure is the right choice of the Cherenkov light lateral distribution function (LDF) suitable for every kind of approximation. Up to now various approximations were only proposed  for  either energies  greater than 1000 TeV \cite{10, 11} (non imaging technique) or up to 10 TeV for the imaging telescopes. In the HiSCORE experiment the method of event reconstruction was primarily developed for the  Tunka-133 wide angle Cherenkov array \cite{12} at energies  above the PeV-range. A special piecewise fitting function \cite{9, 12, 13} (referred below as the ‘Tunka fit’), composed of 4 pieces, was designed to fit both smooth and sharp LDF. The total number of parameters of all constituent functions was reduced to two: $a$ and $bxy$, the first one is a normalization factor, the second one characterizes the steepness of the whole LDF. The Tunka-133 approximation describes all the LDF for primary cosmic rays in the energy interval $10^{15}-10^{18}$~eV, for which it was elaborated. Nonetheless, our analysis reveals that the LDF of gamma rays, especially of ones incident at large zenith angles, very often cannot be reproduced. In \cite{14} we proposed to use a simple 'knee-like' approximation of the Cherenkov light distribution, and have tested the quality of these approximations for gamma rays and hadrons in the energy region of  interest, 30-3000 TeV. In this paper we mainly study the possibility of applying these functions to  real experimental conditions of the HiSCORE experiment.

\section{Event parameters reconstruction in the HiSCORE experiment}
\label{sec-2}
The HiSCORE array  now includes 28 stations spaced with a step of 106 m and covering an area of  0.25 km$^2$. At present the main steps of event reconstruction are the following \cite{9}: 
\begin{enumerate}[1)]
\item Selection of events with more than 4--5 triggered stations (the stations, where the Cherenkov light pulse exceeds the night sky light background). 
\item Rough estimation of zenith and azimuth angles of arriving particles ($\theta,\phi$) using the measured time of delay in every station with a plane approximation of the time front.
\item Reconstruction of the shower core position ($X_0, Y_0$) by minimizing the difference between the experimental $Q_i(R_i)$ and the ‘Tunka fit’, where $R_i$ is the distance between the $i$-th station and the core position ($X_0, Y_0$).
\item Re-estimation of $\theta$ and $\phi$ with the known core position ($X_0, Y_0$) using a more appropriate cone-like fitting function for the time front. It gives significant improvement of the directional resolution from $d\theta{\sim}$1$^\circ$--2$^\circ$ to $d\theta{\sim}$0.1$^\circ$--0.4$^\circ$, because the accuracy of the arrival direction reconstruction depends linearly on that of the shower core position. This point is very important for improving the noise/background ratio of a signal because a background flux depends on the observational solid angle as ${\sim}d\theta^2$. 
\item Approximation of $Q_i(R_i)$ by the 'Tunka fit’ function and estimation of the density of photons at a distance 200 m from the shower core, $Q_{200}$, and corresponding energy estimation by the formula:
\begin{equation}
E=aQ_{200}^{0.94}+b
\end{equation}
($a$ and $b$ are constants obtained from simulations).
\item Parametric analysis of fitting functions for the estimation of the nature of the primary particle.
\end{enumerate}
As is seen, the LDF fitting functions are used on 4 different steps: shower core reconstruction (3), arrival direction estimation (4),  energy estimation (5), and primary particle type identification (6).

\section{Knee-like approximation of the lateral distribution function}
\label{sec-3}

For parameterization of the lateral distribution function, $Q(R)$, of simulated Cherenkov light we used the function that we called a ‘knee-like approximation’, which was used earlier by J. Horandel \cite{15} to  describe the knee in the cosmic ray spectrum. It was a function of energy at that time, but we make it a function of the distance, $R$, from the shower axis. It depends on five parameters $C$, $\gamma_1$, $\gamma_2$, $R_0$, and $\alpha$: 
\begin{equation}
\label{function-1}
F_{appr.} = CR^{\gamma _1}(1+(\frac{R}{R_0})^\alpha)^\frac{\gamma_2}{\alpha}
\end{equation}

In (\ref{function-1}) $R$ is the distance from the shower axis, $R_0$ is the knee position, $\gamma_1$ is the slope of the LDF before the knee, ($\gamma_2$+$\gamma_1$) is the slope of the LDF after the knee, and parameter $\alpha$ characterizes the sharpness of the knee. We applied that fitting function to simulated data to check the quality of the new approximation and used the HiSCORE bank of simulated events (from CORSIKA \cite{16} code), where LDF functions are simulated with a space step of 5 m, for different types of primary particles such as protons, gammas, He, C, Fe in the wide energy interval 30-3000 TeV and zenith angle range 0$^\circ$-50$^\circ$. Below we call them ‘true’ LDF. We showed \cite{14} that the new approximation gives the possibility to fit the whole diversity of individual LDF for different nuclei and gamma rays at  shower core distances of 20–500 m in the energy interval 30–3000 TeV with a very good accuracy. Parameters of the approximation depend on the energy and the type of primary particle and allow us to separate proton and gamma ray induced showers (details will be published in \cite{17}).

However, the 5-parameter fitting function may not be suitable for the real conditions of registration, when we select events with a small number of triggered stations (4--5) to decrease the energy threshold, especially for the task with two additional unknown parameters ($X$-core, $Y$-core). Our study shows that the two-parameter fitting has a distinctive advantage when we work with a small number of triggered detectors; however, it reduces the diversity of LDF to a fixed set of curves. Therefore, different steps of reconstruction require fitting with different number of parameters.

\begin{figure}%[h]
%\centering
\includegraphics[height=5cm,clip]{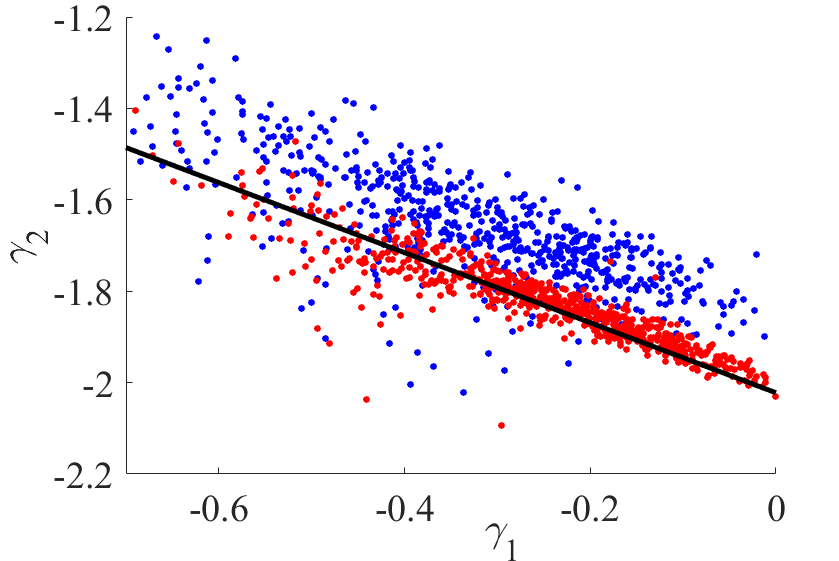}
\caption{Correlation between $\gamma_1$ and $\gamma_2$ parameters in the fitting function (\ref{function-1}) for gamma rays (red points) and protons (blue points) at 100 TeV.}
\label{fig-1}% Give a unique label
\end{figure}

To decrease the number of fitting parameters correlations between parameters were investigated. One of the examples of a strong correlation between $\gamma_1$ and $\gamma_2$ is presented in Fig. \ref{fig-1}. A number of other correlations, including very important correlations of $\gamma_1$, $R_0$, and $\alpha$ with a depth of shower maximum, was also established. Using correlations of different parameters we developed four and three parameter versions of the knee-like fit. The 4-parameter version includes the least squares estimation of only 4 fit parameters: $C$, $\gamma_1$, $R_0$, and $\alpha$, while the fifth parameter, $\gamma_2$, is to be obtained from $\gamma_1$ by an equation:
\begin{equation}
\gamma_2=-0.77\gamma_1-2
\end{equation}
The 3-parameter version sets $\alpha$ to the mean value for the given class of events.

\section{Shower core reconstruction}
\label{sec-4}
\begin{figure}
\centering
%\includegraphics[width=8.5cm,clip]{fig2_end.jpg}
\includegraphics[height=5cm,clip]{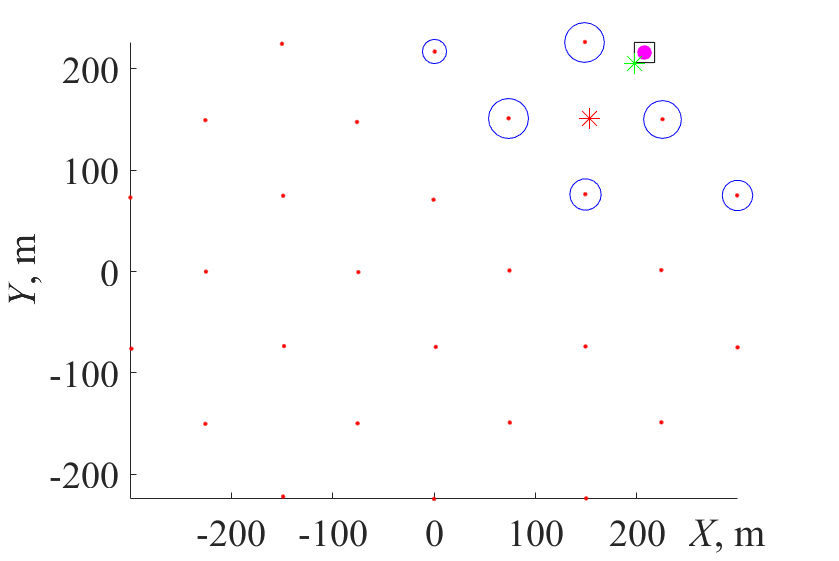}
\includegraphics[height=5cm,clip]{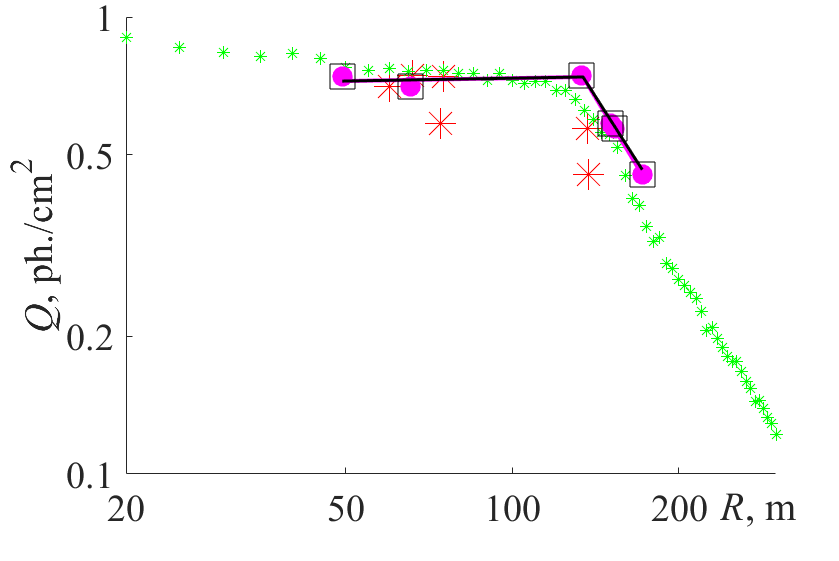}
%\vspace*{10cm}       % Give the correct figure height in cm
\caption{Example of a 100 TeV gamma ray event detected in the HiSCORE array (simulation). Top panel: grid points -- 28 stations of the HiSCORE array, blue circles -- hit detectors with diameter proportional to the value of $Q_i$ in this detector, green star -- true core, red star -- weighted core, purple circle -- core, obtained with 3-parameter fitting, square~--- core, obtained with 4-parameter fitting. Bottom panel: $Q(R)$, where $R$ is calculated from different core positions: red stars -- from the weighted core, purple circles -- from the core, obtained with 3-parameter fitting, squares -- from the core, obtained with 4-parameter fitting. Black line -- 4-parameter fit, purple line -- 3-parameter fit. Green circles -- ‘true’ LDF.}
\label{fig-2}% Give a unique label
\end{figure}

We applied our 3- and 4-parameter fitting function to the step 3 of the  HiSCORE method of event parameters reconstruction. In Fig. \ref{fig-2} we present, for illustration, an individual event from a 100 TeV  gamma ray  detected by 6 stations. In the top panel the stations location is shown and three versions of core position are depicted: the true one, the one obtained by the center of mass technique, and the one found by our method. In the bottom panel the ‘true’ LDF for this event is plotted together with the 3-parameter fitting function. For this event a simultaneous reconstruction of the core position and the shape of the true LDF looks quite satisfactory.

Below we present the statistical analysis of a core position resolution. For that purpose instead of the lateral distribution function,  $Q(R)$, we applied the fit to the so called 'amplitude-distance function',  $A(R)$ \cite{9}, which is the maximal value of the  Cherenkov pulse rather than the total number of photons in the pulse. This variable is more robust and less disturbed by the noise, so it leads to better results in the core position estimation. Besides, the use of $A$ instead of $Q$ allows us to compare both techniques, the 'Tunka fit' and knee-like approximation, on the same kind of data, because the former is applied to $A(R)$ \cite{9}.

\begin{figure}
\centering
%\sidecaption
\includegraphics[height=5cm,clip]{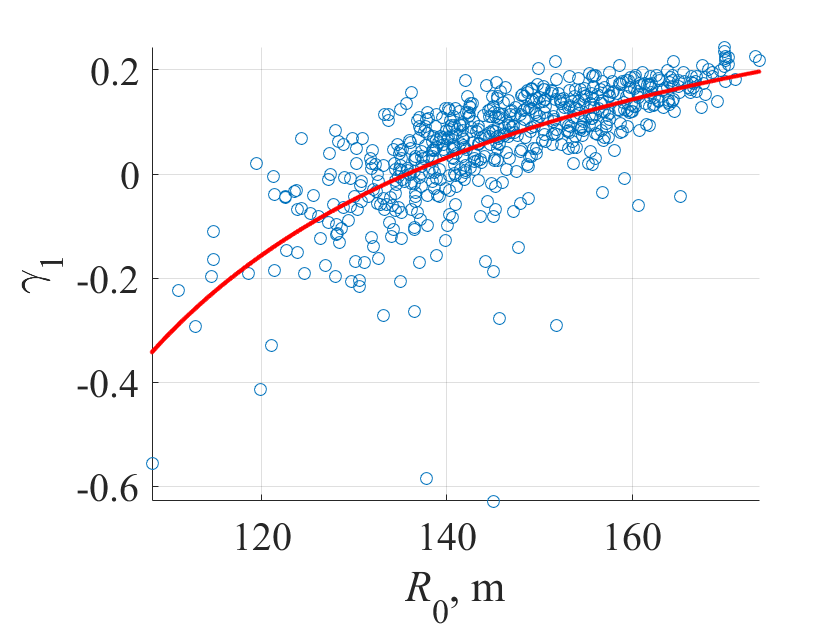}
\includegraphics[height=5cm,clip]{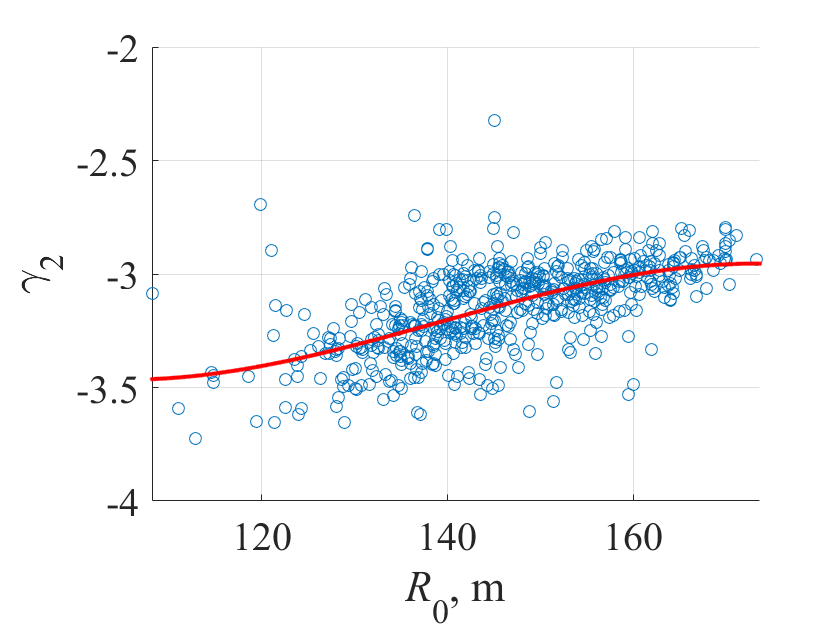}
\caption{Correlation between parameters in the fit (\ref{function-1}) of the amplitude-distance function: $\gamma_1$ and $R_0$ (top panel) and $\gamma_2$ and $R_0$ (bottom panel). Gamma rays at 30-3000 TeV and 28-42$^\circ$}
\label{fig-3}% Give a unique label
\end{figure}

As we noted in section \ref{sec-3}, the 2-parameter method was found to give the best accuracy for the small number of hit detectors. Therefore, we calculated the averaged distance between the fitted core and the true core position for every number of hit detectors using this technique. The two parameters to be fitted to $A(R)$ by the least squares method are $C$ and $R_0$. The other two parameters, $\gamma_1$ and $\gamma_2$, are obtained as functions of $R_0$ using the correlations between them, which were derived from the simulation (Fig. \ref{fig-3}). The values of $\gamma_1$ and $\gamma_2$ as well as the correlation between them differ from those for $Q(R)$.

\begin{figure}
\centering
%\sidecaption
\includegraphics[height=5cm,clip]{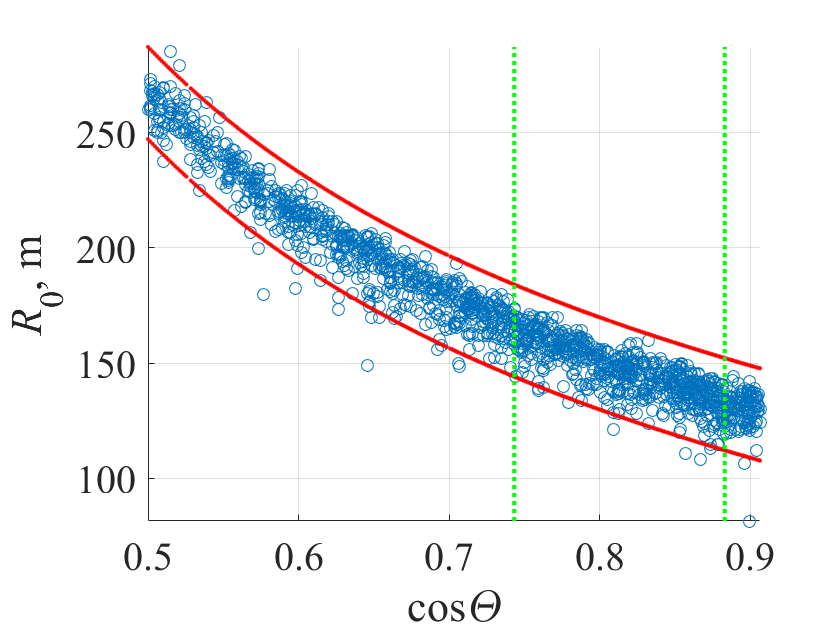}
\includegraphics[height=5cm,clip]{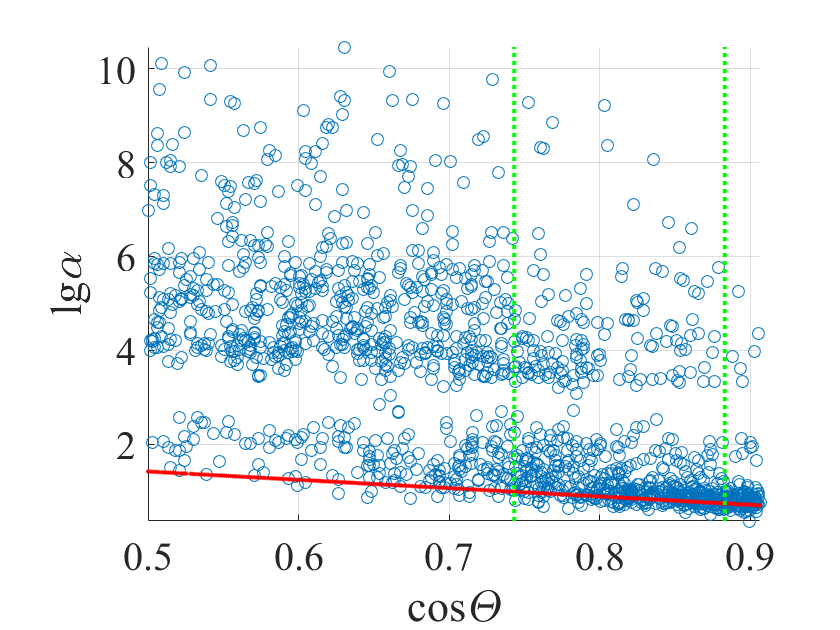}
\caption{$R_0$ limits (top panel) and lg$\alpha$ value (bottom panel) for  use in the 2-parameter knee-like fit of $A(R)$ for  Gamma rays at 30-3000 TeV and 25$^\circ$-60$^\circ$. The dashed vertical lines are to select the angle interval 28-42$^\circ$ for the Crab Nebula observations.}
\label{fig-4}% Give a unique label
\end{figure}

For the best results, the fit parameter $R_0$ is looked for only within the limits (constrained minimization of the mean squared error of the fit), and the limits are made dependent on the shower angle, since the rough estimation of the angle is already performed on the previous stage (step 2 of  section \ref{sec-2}). These limits are derived from the simulations and shown in Fig. \ref{fig-4} as well as the values of the knee sharpness parameter, $\alpha$, which were also found to be slightly dependent on the shower angle. 	

\begin{figure}
\centering
%\sidecaption
\includegraphics[height=5cm,clip]{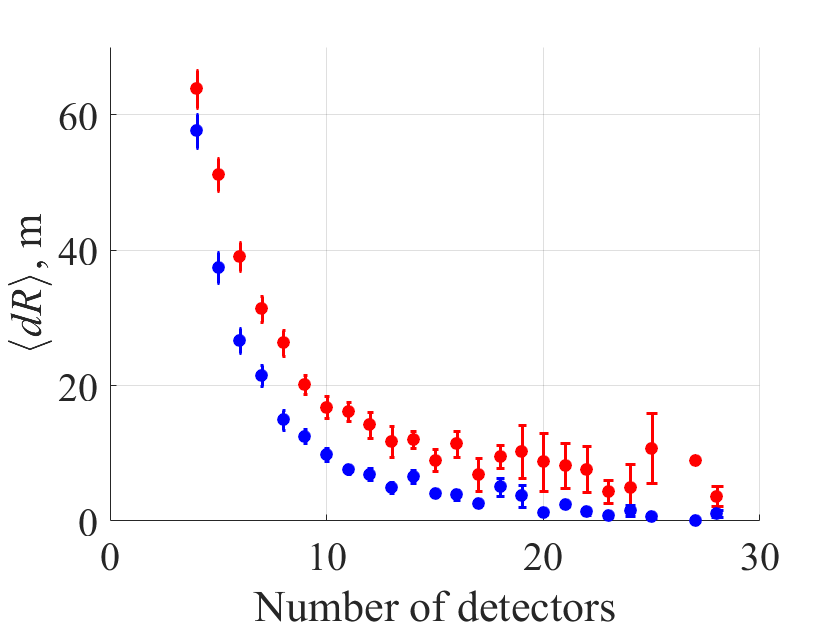}
\caption{The averaged distance between the core position estimate and the true one for gamma events at (30-3000) TeV and 28$^\circ$-42$^\circ$, corresponding to the real conditions of the HiSCORE observation of the Crab Nebula signal. Blue points -- knee-like fit of this article, red points -- Tunka fit.}
\label{fig-5}% Give a unique label
\end{figure}

Finally, in Fig. \ref{fig-5} we plot the mean accuracy ${\langle}dR{\rangle}$ of the core position determined by the 2-parameter version of the knee-like fit, which gives better accuracy for a small number of hit detectors. The figure also contains the results of the `Tunka fit'. The averaging was performed for simulated gamma rays and not for cosmic rays (protons or nuclei), whereas the `Tunka' fit was originally intended for cosmic ray analysis, and that is the reason for the difference in the accuracy of both techniques.

\section{Conclusion}
\label{sec-5}
We proposed and studied a 'knee-like' 5-parameter approximation of lateral distributions of Cherenkov light emitted by extensive air showers. The knee-like approximation is capable of describing the whole diversity of individual lateral distribution functions for different nuclei and gamma rays at the shower core distance 20-500 m in the energy interval (30-3000) TeV with  very good accuracy. Using correlations between different parameters of the knee-like fit we can decrease the number of parameters to 4, 3, or 2. Good quality of these versions of the knee-like fit was tested on simulated data in conditions of the real experimental setup of the TAIGA-HiSCORE project. An algorithm of a shower core reconstruction based on this fit was developed and tested on the same data.

This research confirms an ability of the knee-like fit technique to work with  real data and its potential in improving the accuracy of event reconstruction in wide angle non-imaging experiments.

The study was supported by the Russian Foundation for Basic Research, project no. 16-29-13035.

% BibTeX or Biber users please use (the style is already called in the class, ensure that the "woc.bst" style is in your local directory)
% \bibliography{name or your bibliography database}

\end{document}